\documentclass[conference]{IEEEtran}
\IEEEoverridecommandlockouts
\usepackage[dvipsnames]{xcolor}
\usepackage[hidelinks,colorlinks=true,urlcolor=Magenta]{hyperref}       %
\usepackage{url}            %
\usepackage[misc,geometry]{ifsym}
\usepackage{cite}
\usepackage{amsmath,amssymb,amsfonts}
\usepackage{algorithmic}
\usepackage{graphicx}
\usepackage{textcomp}
\usepackage{xcolor}
\usepackage{caption}
\usepackage{booktabs}
\usepackage{multirow}

\newcommand\extrafootertext[1]{%
    \bgroup
    \renewcommand\thefootnote{\fnsymbol{footnote}}%
    \renewcommand\thempfootnote{\fnsymbol{mpfootnote}}%
    \footnotetext[0]{
    \vspace{-0.7cm}
    \par#1}%
    \egroup
}

\def\BibTeX{{\rm B\kern-.05em{\sc i\kern-.025em b}\kern-.08em
    T\kern-.1667em\lower.7ex\hbox{E}\kern-.125emX}}
\begin{document}

\title{MedDet: Generative Adversarial Distillation for Efficient Cervical Disc Herniation Detection}

\author{\IEEEauthorblockN{\fontsize{10}{4}\selectfont Zeyu Zhang$^{1}$$^{2}$$^{*}$$^{\dag}$, Nengmin Yi$^{1}$$^{*}$, Shengbo Tan$^{1}$$^{*}$, Ying Cai$^{1}$$^{\text{\Letter}}$, Yi Yang$^{3}$, Lei Xu$^{4}$, 
Qingtai Li$^{1}$, 
Zhang Yi$^{4}$, 
Daji Ergu$^{1}$, 
Yang Zhao$^{5}$}

\IEEEauthorblockA{$^{1}$Key Laboratory of Electronic Information Engineering, Southwest Minzu University\\
$^{2}$The Australian National University
$^{3}$Orthopedic Research Institute, West China Hospital\\
$^{4}$Machine Intelligence Laboratory, Sichuan University
$^{5}$La Trobe University}

\href{https://steve-zeyu-zhang.github.io/MedDet}{https://steve-zeyu-zhang.github.io/MedDet}}

\makeatletter
\let\@oldmaketitle\@maketitle%
\renewcommand{\@maketitle}{\@oldmaketitle%
\includegraphics[width=\textwidth]{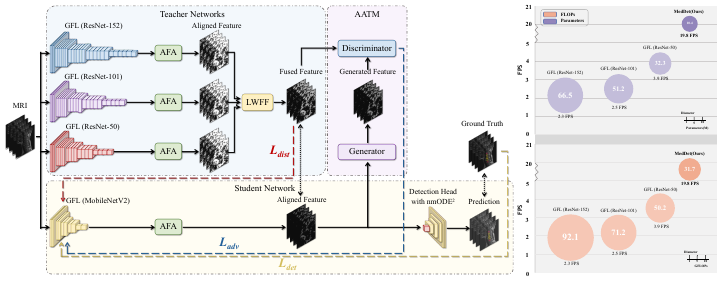}\setcounter{figure}{0}
    \captionof{figure}{The left subfigure illustrates the overall architecture of our proposed MedDet model, which includes a novel adversarial auxiliary teacher module (AATM) for generative adversarial distillation, an adaptive feature alignment (AFA), and a learnable weighted feature fusion (LWFF) module to fuse features from teacher networks and align them with those of the student network. Additionally, it incorporates a denoising nmODE$^2$ module in the detection head. The right subfigure shows that our method achieves superior efficiency compared to teacher models in terms of FLOPs, parameter count, and inference speed.}
    \label{fig:main}\bigskip}%
\makeatother

\maketitle

\extrafootertext{\noindent\rule{0.5\linewidth}{0.4pt}
\\$^{*}$Equal Contribution. $^{\dag}$Work done in Southwest Minzu University.
\\$^{\text{\Letter}}$Corresponding author: \href{mailto:caiying34@yeah.net}{caiying34@yeah.net}}

\begin{abstract}
Cervical disc herniation (CDH) is a prevalent musculoskeletal disorder that significantly impacts health and requires labor-intensive analysis from experts. Despite advancements in automated detection of medical imaging, two significant challenges hinder the real-world application of these methods. First, the computational complexity and resource demands present a significant gap for real-time application. Second, noise in MRI reduces the effectiveness of existing methods by distorting feature extraction. To address these challenges, we propose three key contributions: Firstly, we introduced MedDet, which leverages the multi-teacher single-student knowledge distillation for model compression and efficiency, meanwhile integrating generative adversarial training to enhance performance. Additionally, we customize the second-order nmODE to improve the model's resistance to noise in MRI. Lastly, we conducted comprehensive experiments on the CDH-1848 dataset, achieving up to a 5\% improvement in mAP compared to previous methods. Our approach also delivers over 5 times faster inference speed, with approximately 67.8\% reduction in parameters and 36.9\% reduction in FLOPs compared to the teacher model. These advancements significantly enhance the performance and efficiency of automated CDH detection, demonstrating promising potential for future application in clinical practice.
\end{abstract}


\section{Introduction}

Neck pain is a common musculoskeletal disorder and is the fourth leading cause of disability, presenting a significant public health challenge and burdening patients and healthcare systems \cite{cohen2017advances,cohen2015epidemiology}. The \textit{Global Burden of Disease Study 2017} reported that neck pain affects 288.7 million people, with a notable increase over the past three decades \cite{safiri2020global}. A common cause of neck pain is the degeneration of cervical intervertebral discs \cite{risbud2014role,fujimoto2012sensory,theodore2020degenerative}. This degenerative process can result in cervical disc herniation (CDH), where fragments of the disc, particularly the nucleus pulposus, protrude into the spinal canal through a ruptured annulus fibrosus\cite{takahashi2013laterality}. This herniation can compress the spinal cord or nerve roots, leading to various clinical symptoms.

T2-weighted magnetic resonance imaging (MRI) is the gold standard for diagnosing CDH due to its superior ability to detect disc morphology and the hydration status of the nucleus pulposus \cite{farshad2015mr,almansour2022deep}. Despite these advantages, diagnosing CDH remains challenging and depends heavily on the expertise of radiologists or surgeons, who must undergo extensive training to accurately interpret and analyze biomedical images. Although experienced orthopedic surgeons can visually inspect MRIs to identify CDH features, the high workload can cause fatigue and pressure, highlighting the need for an accurate and rapid automatic detection method.

Medical imaging analysis \cite{zhang2024jointvit,wu2024xlip,hiwase2024can,zhao2024landmark} has made significant advancements across various dense prediction tasks \cite{ge2024esa}, particularly in medical image detection and segmentation \cite{wu2023bhsd,zhang2024segreg,tan2024segstitch,zhang2023thinthick}. These improvements have enhanced the ability to accurately identify and delineate anatomical structures, abnormalities, and disease markers, which ultimately contributes to better diagnosis, treatment planning, and patient outcomes \cite{zhang2024deep}. Therefore, developing robust and effective methods for detecting CDH appears to be a highly promising direction for future research. However, creating a practical CDH detection method that benefits healthcare currently faces two significant challenges. 

Firstly, given the high volume of MR imaging and the limited capabilities of medical equipment, real-time detection of CDH is crucial for meeting clinical needs. This leads to an unavoidable trade-off between model performance and efficiency when considering real-world applications.

Secondly, noise in MRI images cannot be ignored. Factors like patient movement, metallic implants, and external electromagnetic waves during MRI scans can degrade image quality, resulting in unclear data and alter the contour, color, and position of intervertebral discs, thereby complicating the diagnostic process. Therefore, developing a method that effectively handles MRI noise without distorting feature extraction is quite challenging.

However, knowledge distillation in object detection has been widely adopted for improving model efficiency in general computer vision due to its ability to distill high-semantic features for boosting lightweight models. Meanwhile, generative adversarial learning has been increasingly adapted to enhance distilled features for improving student model's performance. Therefore, customizing knowledge distillation and generative adversarial training represents a promising direction for advancing CDH detection.

Besides, Ordinary differential equations (ODEs) provide a mathematical foundation for studying dynamic systems, and biological neural networks can be viewed as such systems. This perspective has led to a growing interest in integrating ODEs into deep neural networks \cite{weinan2017proposal,chang2019antisymmetricrnn}. Unlike traditional neural networks that use discrete neurons and layers, ODE-based models can represent the dynamic processes of neural networks as continuous-time systems. Describing neuron activity and network behavior using differential equations enhances our understanding of the dynamic properties of neural networks, allowing us to study their performance across various tasks and environments. Moreover, employing differential equation-solving methods to train and optimize neural networks enables the incorporation of the system's dynamic characteristics into the training process, thereby improving both performance and resilience to noise in MRI.

To overcome these challenges, our paper introduces three key contributions:

\begin{itemize}
    \item Firstly, we introduce \textbf{MedDet}, a novel framework that combines multi-teacher single-student knowledge distillation to achieve model compression and efficiency. This approach allows the model to maintain high performance while reducing computational costs, making it more suitable for real-world medical applications. Additionally, we introduce an \textbf{adversarial auxiliary teacher module} (\textbf{AATM}) to enhance the model's accuracy and robustness by improving its generalization across diverse medical imaging scenarios. We also design an \textbf{adaptive feature alignment} (\textbf{AFA}) and \textbf{learnable weighted feature fusion} (\textbf{LWFF}) to dynamically align and fuse the features of teacher models during distillation.
    \item Additionally, we customize \textbf{nmODE$^2$} to enhance the model's resistance to noise in MRI scans. This advanced architectural component is specifically designed to address the challenges posed by noisy MRI data, such as those resulting from patient movement or external interference. By integrating the nmODE$^2$, our model significantly improves its ability to extract accurate image representations, even in the presence of noise, thereby making the diagnostic process more reliable and robust in clinical settings.
    \item Lastly, We conducted comprehensive experiments on CDH-1848 dataset, where our approach demonstrated up to a \textbf{5\%} improvement in mAP compared to previous methods. Furthermore, our model achieved an approximate \textbf{67.8\%} reduction in parameters and a \textbf{36.9\%} reduction in FLOPs compared to the teacher model. These advancements not only boost the performance of automated CDH detection but also enhance its efficiency, making it more practical for real-world clinical applications where both accuracy and computational resource optimization are crucial.
\end{itemize}

\section{Related Works}

\subsection{Traditional Methods} 
Ghosh et al. \cite{ghosh2011computer} developed a lumbar disc herniation detection system that integrated a mathematical model and machine learning to identify ROIs using probability and active shape models, followed by classification with five models via majority voting. However, the system was manual-intensive, inefficient in feature extraction, and time-consuming.
Koh et al.\cite{koh2012disc} presented a method using diverse classifiers, yet deemed unfit for clinical use. 
Unal et al. \cite{unal2015detection} proposed a hybrid methods for abnormal intervertebral disc detection, involving feature extraction, selection, and classification. Yet, reliance on manual feature extraction by doctors hindered full automation. 
Ebrahimzadeh et al. \cite{ebrahimzadeh2018machine} presented a lumbar disc herniation diagnosis method that integrates various algorithms, employing Otsu's thresholding and machine learning technologies. Yet, the feature extraction process was complex and challenging to meet real-time requirements.
Hashia et al. \cite{hashia2020texture} proposed a method to classify normal and herniated intervertebral discs in the sagittal plane by extracting ROIs from MRI images, applying a gray-level run-length matrix for texture analysis, and employing a BPNN classifier for the final classification.

\subsection{Deep Learning Methods} 
In recent years, deep learning has seen rapid development, gradually supplanting traditional machine learning techniques as the predominant approach for CDH detection. 
For instance, Pan et al.\cite{pan2021automatically} 
introduced a step-by-step CNN model based on the two-stage detection network faster-RCNN, categorizing intervertebral discs into normal, bulging, and protruding types. However, this algorithm does not support end-to-end processing, has a complex implementation, involves large model parameters, and exhibits slow inference speeds, which are not conducive to edge device deployment.
Similarly, Tsai et al. \cite{tsai2021lumbar}
proposed a data augmentation method for lumbar intervertebral disc protrusion detection using the one-stage detection network YOLOv3 \cite{redmon2018yolov3}, which, despite improving the network's generalization, falls short in meeting clinical diagnostic accuracy requirements. Chen et al. \cite{chen2021usage} proposed a deep learning-based intelligent auxiliary diagnosis system. The system uses an improved activation function TReLU \cite{nakhua2023trelu} to construct and optimize the CDCGAN network model and establishes an auxiliary diagnosis system including MRI feature extraction, 3D reconstruction, and CDCGAN classification. Guinebert et al. \cite{guinebert2022automatic} proposed a semi-automatic system for segmented lumbar intervertebral disc disease diagnosis. While these methods address aspects of timeliness or accuracy for intervertebral disc protrusion tasks, there is a clinical need for a balance between efficiency and performance.

\section{Methodology}

\subsection{Overview}
In this paper, we introduce a novel method based on generative adversarial knowledge distillation and design an innovative feature alignment and fusion method, along with a new denoising block, \textbf{nmODE$^2$}, to enhance the detection of cervical intervertebral disc herniation. Our approach is designed to deliver high performance while maintaining a balanced and efficient parameter scale, ensuring both accuracy and practicality in clinical applications. Firstly, we introduce an \textbf{adversarial auxiliary teacher module} (\textbf{AATM}) that utilizes three teacher networks and one student network, incorporating a feature pyramid network (FPN) \cite{lin2017feature} to improve object detection across different scales. The networks are trained with generalized focal loss (GFL) \cite{li2020generalized} for CDH detection and a second-order ordinary differential neural network (nmODE$^2$) to address MRI noise and enhance network robustness. The teacher networks use ResNet-50, ResNet-101, and ResNet-152 \cite{he2016deep} as their backbones, while the student network employs MobileNetV2 \cite{sandler2018mobilenetv2} as its backbone, as shown in Fig. \ref{fig:main}. All teacher networks are pretrained on our CDH dataset. We then customize \textbf{adaptive feature alignment} (\textbf{AFA}) and \textbf{learnable weighted feature fusion} (\textbf{LWFF}) to align the student network’s features with those of the teacher networks. During knowledge distillation, we incorporate adversarial training in AATM to effectively transfer knowledge from the teacher networks to the student network, ensuring stable and efficient feature transfer. Finally, the output features from the student network are fed into a detection head to perform box regression and classify the cervical disc.

\subsection{Generative Adversarial Knowledge Distillation}

\begin{figure}[htbp]
\centering
\includegraphics[width=\linewidth]{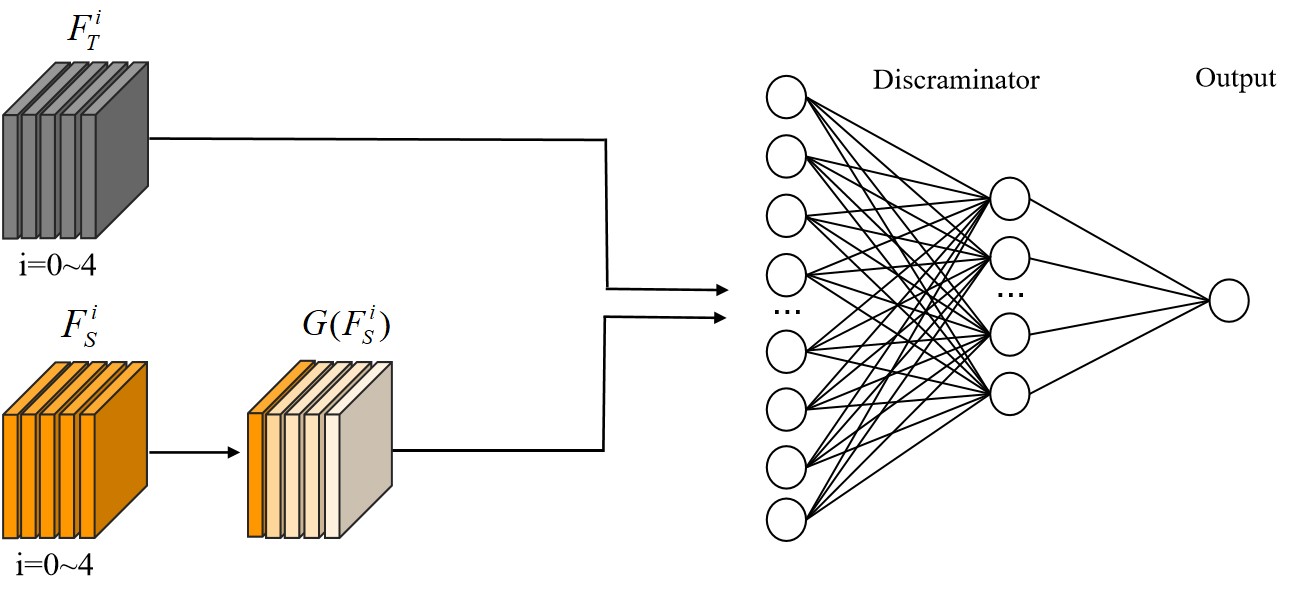}
\caption{The diagram of the adversarial auxiliary teacher module (AATM). The $F_{T}^{i}$ and $F_{S}^{i}$ represent the \textit{i}-th feature maps outputted by the teacher network and the student network's FPN respectively. \textit{G} represents the generator.}
\label{fig:gan}
\end{figure}

To enhance the effectiveness of distillation, we incorporate an adversarial mechanism to ensure robustness. We introduce an adversarial auxiliary teacher module (AATM) that guides the student network to consistently restore the expressive features of the teacher network. Inspired by GANs \cite{goodfellow2020generative}, we use features extracted by the teacher network as positive samples and those generated by the student network as negative samples. AATM assists the student network in restoring the diverse feature expressions of the teacher network.

As shown in Figure \ref{fig:gan}, the discriminator distinguishes between positive and negative samples, and its outputs are used to calculate the losses for both the generator and the student network, enabling parameter updates. This process encourages the student network's output features to closely resemble those of the teacher network. By leveraging the discriminator to differentiate between positive and negative samples, the student network learns more effectively from the teacher network, and the generator progressively approaches ground truth images through iterative refinement.

The generator is defined with two 3x3 convolutional layers and an activation layer to enhance pixel correlation, while the discriminator is defined with three fully connected layers and an activation layer to classify positive and negative samples. Binary cross-entropy loss is used to measure the similarity between positive and negative samples, supervising the training of the discriminator network. During training, the teacher network’s parameters are frozen, and the intermediate features of the student network are trained through adversarial learning. The discriminator maximizes the \textit{D} loss to determine whether an image originates from the teacher or student network. The training process of AATM can be represented as follows:

\begin{equation}\label{11}
    \begin{split}
    \underset{G}{min}\;\underset{D}{max}V\left ( D,G\right ) &=\sum_{i=0}^{i=4}(\mathbb{E}_{{F_{T}^{i}}\sim{P_{T}^{i}}}\left [ \log D\left ( F_{T}^i\right )\right ] + \\&\quad 
     \mathbb{E}_{{F_{S}^i}\sim{P_{S}^i}}\left [\log\left (1- D\left ( G\left ( F_{S}^i\right )\right )\right )\right ])
    \end{split}
\end{equation}

In the equation, $F_{T}$ and $F_{S}$ denote the intermediate output features of the teacher and student networks, respectively, while ${P_{T}}$ and ${P_{S}}$ represent the original distributions of these intermediate features. The index $i$ is used for generating the intermediate output feature for discrimination. \textit{D} and \textit{G} represent the discriminator and generator, respectively.

\subsection{The nmODE$^2$ Block}

\begin{figure}[htbp]
    \centering 
    \includegraphics[width=\linewidth]{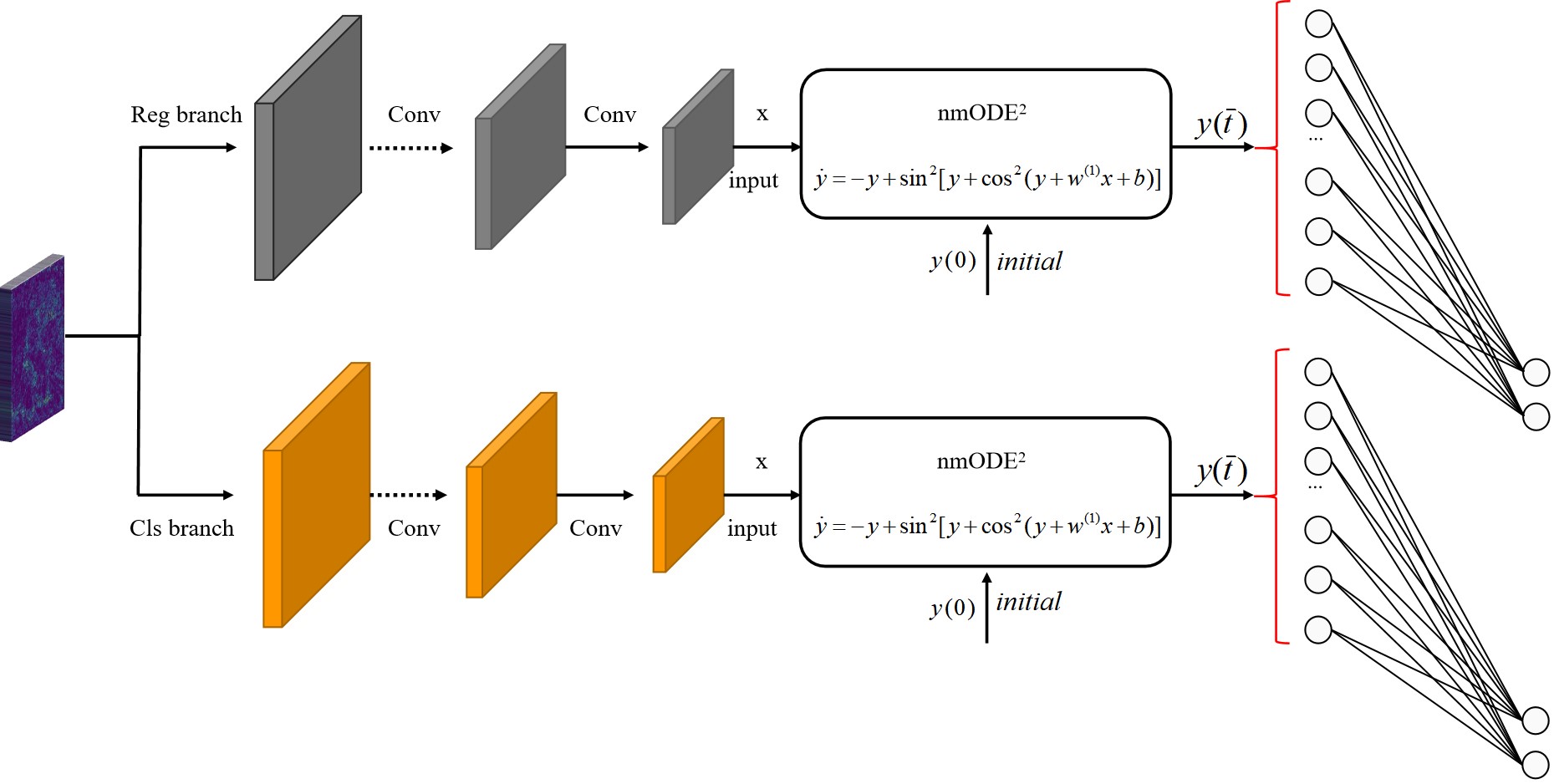}
    \caption{The diagram illustrates the architecture of the nmODE$^2$ block, integrated into the classification and regression heads of the teacher networks.}
    \label{fig:nmode}
\end{figure}

The nmODE \cite{yi2023nmode} is a recently introduced approach for solving neural networks using ordinary differential equations, which can be expressed as

\begin{equation}
    \dot{y} = -y + \sin^{2} (y+\gamma )
    \label{7}
\end{equation}
\begin{equation}
    \gamma = wx + b
    \label{8}
\end{equation}

In these equations, $x$ represents the input feature map, $y$ denotes the network state, $\gamma$ indicates the perceptual input, $w$ signifies the learning weight, $b$ stands for the bias, and $\sin$ represents the activation process.

In this work, we introduce nmODE$^{2}$, an extension of nmODE \cite{yi2023nmode}, to optimize the neural network model. The nmODE$^{2}$ has well-defined dynamics and global attractor properties. It introduces a memory mechanism that utilizes attractors within the network to establish a relationship between external inputs and memory. Like nmODE, nmODE$^{2}$ features a unique structure that separates learning and memory functions. Each learning connection exists only between one input neuron and one memory neuron, with no connections between memory neurons. Additionally, nmODE$^{2}$ has strong nonlinear modeling capabilities, enhancing the network's depth and ability to model complex relationships.

We integrate nmODE$^{2}$ into the classification and regression heads of the teacher networks, as shown in Fig. \ref{fig:nmode}. By leveraging nmODE$^{2}$'s distinct separation of memory and learning functions, we can effectively learn the features of cervical intervertebral discs by inputting them into the teacher network's classification and regression heads, then storing the learned features in the memory module. This approach allows the network to retain the morphological characteristics of cervical intervertebral discs and significantly improves its generalization ability in noisy images, thereby enhancing its detection capability for CDH and reducing the impact of noise.

For any given input feature map $x$, the $nmODE^2$ produces a unique global output $y(\bar{t})$. Thus, $nmODE^2$ defines a nonlinear mapping from the external input feature map $x$ to the output $y(\bar{t})$, which can be expressed as:

\begin{equation}
    \dot{y} = -y + \sin^{2}\left [ y + \cos^{2}\left ( y + \gamma \right )\right ]
\end{equation}
\begin{equation}
    \gamma = f\left (x,w \right )
\end{equation}

In these equations, $x$ represents the input feature map, $y$ denotes the network state, $\gamma$ signifies the perceptual input, $w$ stands for the learning weight, and $b$ represents the bias. The terms $\sin$ and $\cos$ correspond to the activation process, while $f$ represents the mapping relationship from the upstream neural network input to the output.

\subsection{Feature Alignment and Fusion}

The feature spaces of different teacher and student networks are distinct, making the effective fusion and alignment of features from multiple teachers and the student network two critical challenges that directly impact performance. To address these challenges, we propose an adaptive feature alignment (AFA) method to standardize the output feature sizes of three teacher networks and a student network. Following this, we introduce a learnable weighted feature fusion (LWFF) technique to combine the output features of the three teacher networks, facilitating effective adversarial training and knowledge distillation with the student network. However, traditional knowledge distillation methods using a single teacher often struggle to balance parameter reduction with performance stability. To address this, we introduce a multi-teacher single-student distillation framework based on \cite{romero2014fitnets}.

\subsubsection{Adaptive Feature Alignment (AFA)}

\begin{figure}[htbp]
    \centering
    \includegraphics[width=\linewidth]{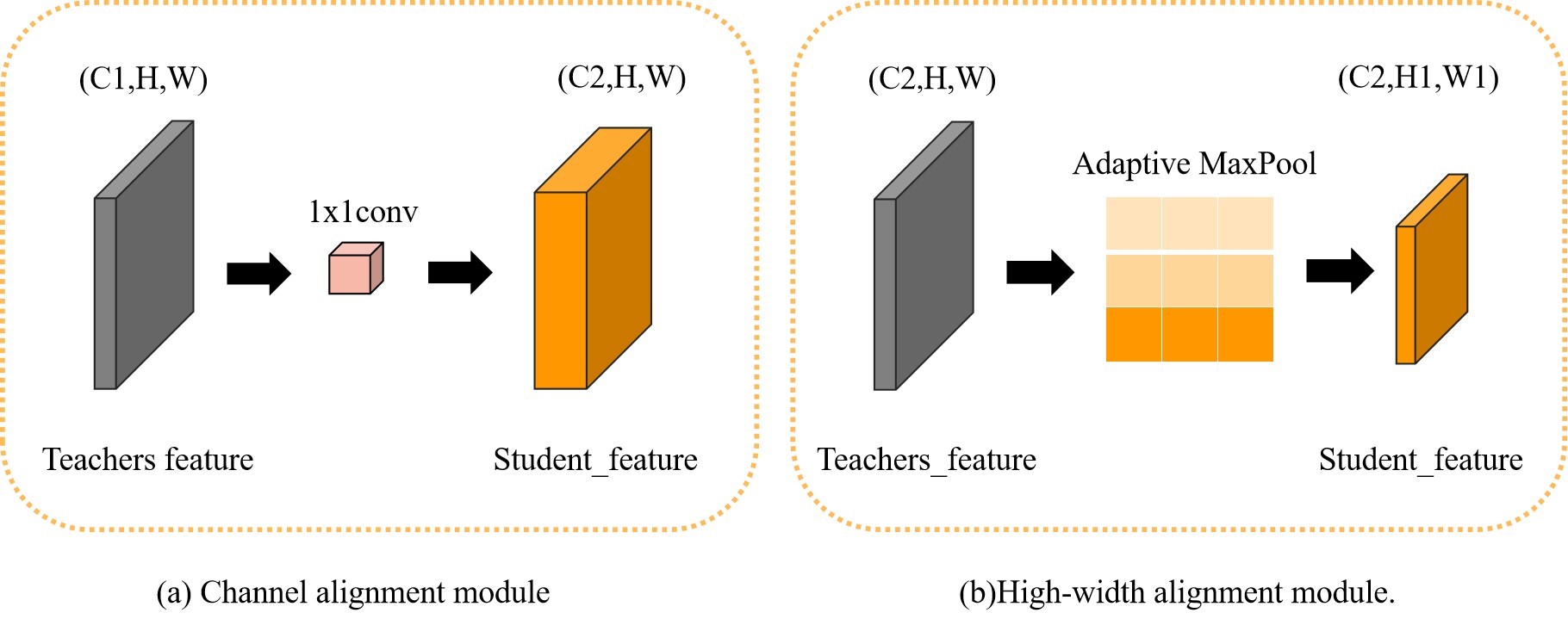}
    \caption{The figure illustrates the adaptive feature alignment (AFA). Subfigure (a) shows the channel-wise alignment model, while subfigure (b) shows the height-width (\textit{HW}) alignment model.}
    \label{fig:adaption_module1_m}
\end{figure}

The adaptive alignment module (AFA) consists of two components: the channel-wise alignment module and the height-width (\textit{HW}) alignment module. First, the channel-wise alignment module adapts a 1 $\times$ 1 convolution operation to align the channels of the output feature maps from the teacher and student networks, as shown in Fig. \ref{fig:adaption_module1_m}(a). Next, the feature maps from the teacher network, processed by the channel alignment module, are passed to the \textit{HW} alignment module. Here, the teacher network's feature maps are resized to match the dimensions of the student network's feature maps using adaptive max pooling, as illustrated in Fig. \ref{fig:adaption_module1_m}(b). The overall process of the adaptive alignment module can be represented as follows:

\begin{equation}\label{1}
\textit{align} = \textit{adp}\left [ \textit{Conv}_{1 \times 1}\left ( F_{T}\right )\right ]
\end{equation}

The \textit{adp} denotes adaptive max pooling, $\textit{Conv}_{1 \times 1}$ refers to a 1 $\times$ 1 convolution, and $F_{T}$ represents the intermediate features of the teacher network.

\subsubsection{Learnable Weighted Feature Fusion (LWFF)}

\begin{figure}[htbp]
\centering
\includegraphics[width=\linewidth]{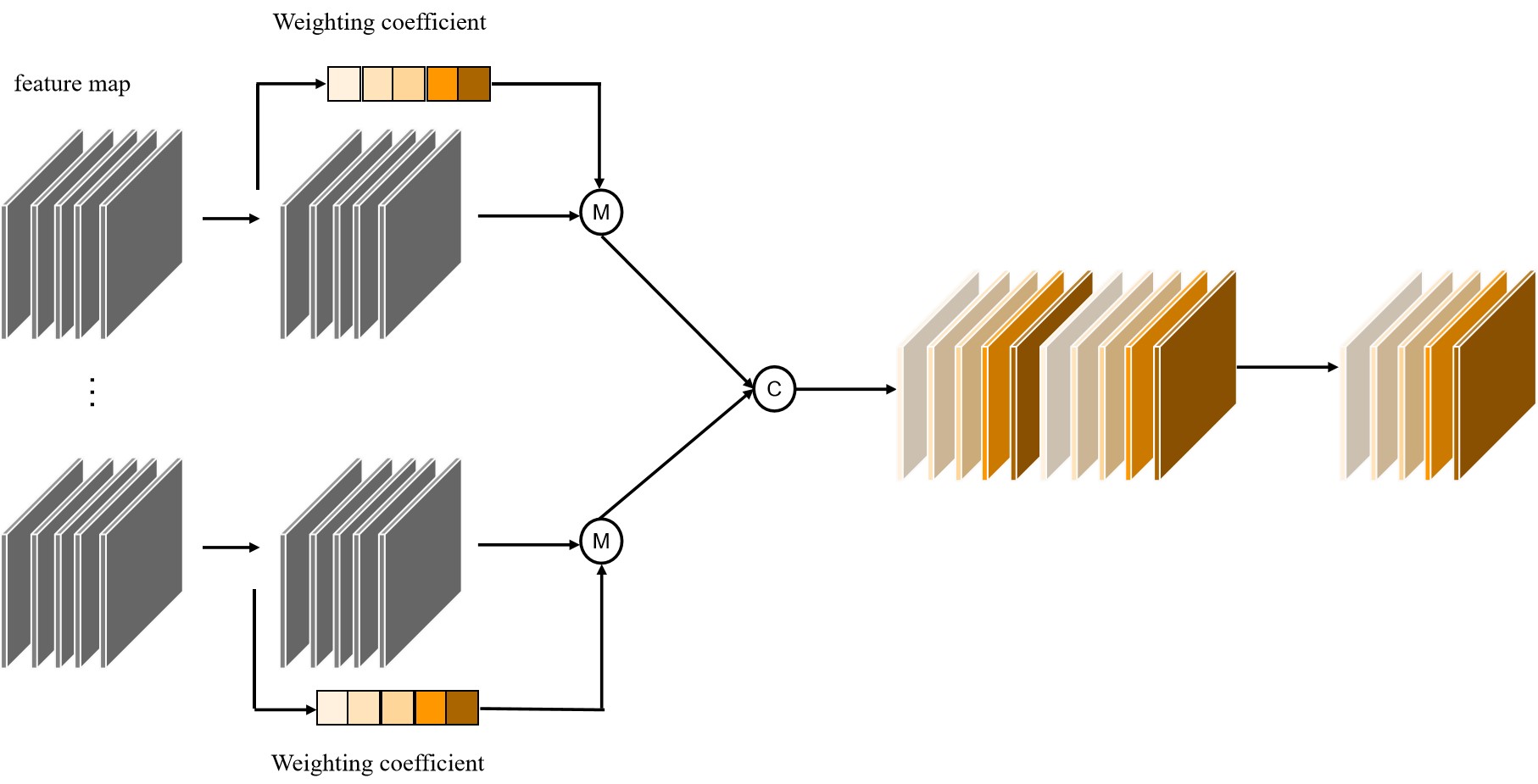}
\caption{The figure illustrates the learnable weighted feature fusion (LWFF) module, where $M$ denotes the multiplication operation and $C$ denotes the concatenation operation.}
\label{fig:fusion_module3}
\end{figure}

In order to effectively integrate the intermediate features of the teacher network, we propose a learnable weighted feature fusion (LWFF) module. This module dynamically assigns weight coefficients to different channel feature maps, allowing for the adaptive acquisition of fusion features that provide greater benefits for the task, as shown in Fig. \ref{fig:fusion_module3}.

Specifically, we first pass the intermediate output features of each teacher network to the LWFF module. We perform a global average pooling operation on each intermediate output of the teacher network and then apply convolutional operations to extract non-linear relationships between different channels. Next, we adaptively assign different weighting coefficients based on the importance of each channel and multiply these coefficients with the original input features to obtain the weighted features for each teacher network. Finally, we concatenate the weighted features and use a 1 $\times$ 1 convolution to extract features and reduce their dimensions.

The intermediate feature outputs of each teacher network are denoted as $F_{T_1}$, $F_{T_2}$, and $F_{T_3}$, where $\alpha$ represents the adaptive weighting module, $\textit{global}$ denotes the global average pooling operation \cite{lin2013network}, $\textit{Conv}_{1 \times 1}$ signifies the 1 $\times$ 1 convolution operation, and $*$ indicates the multiplication operation. The outputs of each teacher network after passing through the adaptive weighting feature module are $F_{T_1}^{'}, F_{T_2}^{'},$ and $F_{T_3}^{'}$. The $\textit{cat}$ operation denotes concatenation, and $F_{\textit{fusion}}$ represents the final output of the LWFF. The specific process can be described as follows:

\begin{equation}
    \alpha = \textit{Conv}_{1 \times 1}\left ( \textit{global}\left ( x\right )\right )
\end{equation}
\begin{equation}
    F_{T_{1}}^{^{'}}= \alpha \left ( F_{T_{1}}\right )*F_{T_{1}}
\end{equation}
\begin{equation}
    F_{T_{2}}^{^{'}}= \alpha \left ( F_{T_{2}}\right )*F_{T_2}
\end{equation}
\begin{equation}
     F_{T_{3}}^{^{'}}= \alpha \left ( F_{T_{3}}\right )*F_{T_3}
\end{equation}
\begin{equation}
    F_{\textit{fusion}}= \textit{Conv}_{1 \times 1}\left [ \textit{cat}\left ( {F_{T_{1}}'},{F_{T_{2}}'},{F_{T_{3}}'}\right )\right ]
\end{equation}

\subsection{Optimization}

We optimize the student model by computing three types of losses: the detection loss of the student network $L_\textit{det}$, the distillation loss $L_\textit{dist}$, and the adversarial loss between the feature maps of the teacher and student networks $L_\textit{adv}$. 

The detection loss $L_\textit{det}$ for the student network is represented as:
\begin{equation}
    L_\textit{det} = \lambda L_\textit{QF} + \mu L_\textit{DF} + (1 - \lambda - \mu) L_\textit{GIoU}
\end{equation}

\noindent where quality focal loss $L_\textit{QF}$ is adapted for the classification, distribution focal loss $L_\textit{DF}$ is adapted for the regression, GIoU loss $L_\textit{GIoU}$ is adapted for the localization, and $\lambda, \mu$ are hyperparameters.

The distillation loss $L_\textit{dist}$ can be represented as:
\begin{equation}
    L_\textit{dist}\left ( S,T\right )=\sum_{i=0}^{i=4}\left ( T_{i}-G\left [ \textit{align}\left ( S_{i}\right )\right ]\right )^{2}
\end{equation}

\noindent where $i$ is the index of the FPN output features for both the teacher and student networks, $T_i$ represents the fused feature map from the teacher network, $S_i$ denotes the feature map from the student network, $\textit{align}$ refers to the feature alignment module, and $G$ represents the generation module.

Therefore, the total loss $L$ is given by:
\begin{equation}
    L = \sigma L_\textit{det} + \tau L_\textit{dist} + (1 - \sigma - \tau) L_\textit{adv}
\end{equation}
\noindent where $\sigma$ and $\tau$ are hyperparameters.

\noindent\textbf{\textit{Why not KL divergence loss?}} \textit{KL} divergence is typically applied to match the output distributions of networks. In contrast, our optimization focuses on ensuring that the intermediate features of the teacher and student networks are similar, rather than aligning probability distributions.

\section{Dataset and Evaluation Matrices}

\subsection{CDH-1848 Dataset}

\begin{figure}[htbp]
    \centering
    \includegraphics[width=\linewidth]{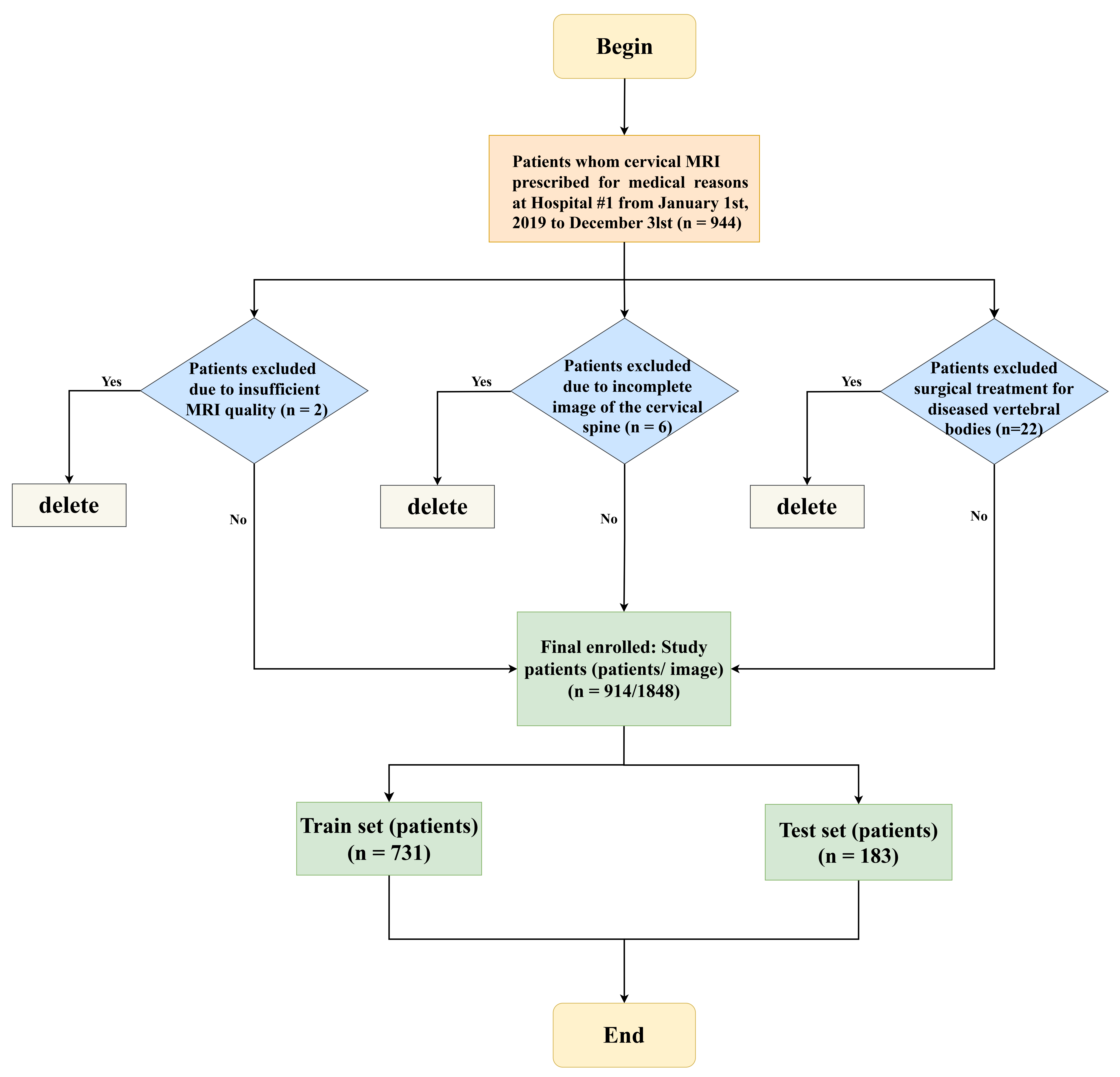}
    \caption{The figure illustrates the pipeline for patient selection, MRI acquisition, and expert annotation following inter-rater reliability.}
    \label{fig:dataset}
\end{figure}

The CDH-1848 dataset consists of 1,848 de-identified MR images from 914 patients, which were annotated by radiologists following inter-rater reliability. The images were categorized into two labels: protrusion and non-protrusion, with corresponding bounding boxes, as shown in Fig. \ref{fig:dataset}.

\subsection{Evaluation Matrices}

We utilized multiple evaluation metrics in our experiments, including mAP and recall. mAP measures the precision of the model in detecting intervertebral discs, averaging the precision across different classes and thresholds to provide an overall performance score. Recall indicates the model's ability to detect all intervertebral discs, highlighting its effectiveness in minimizing false negatives. 

\section{Comparative Studies}

\begin{figure}[htbp]
\centering
\includegraphics[width=\linewidth]{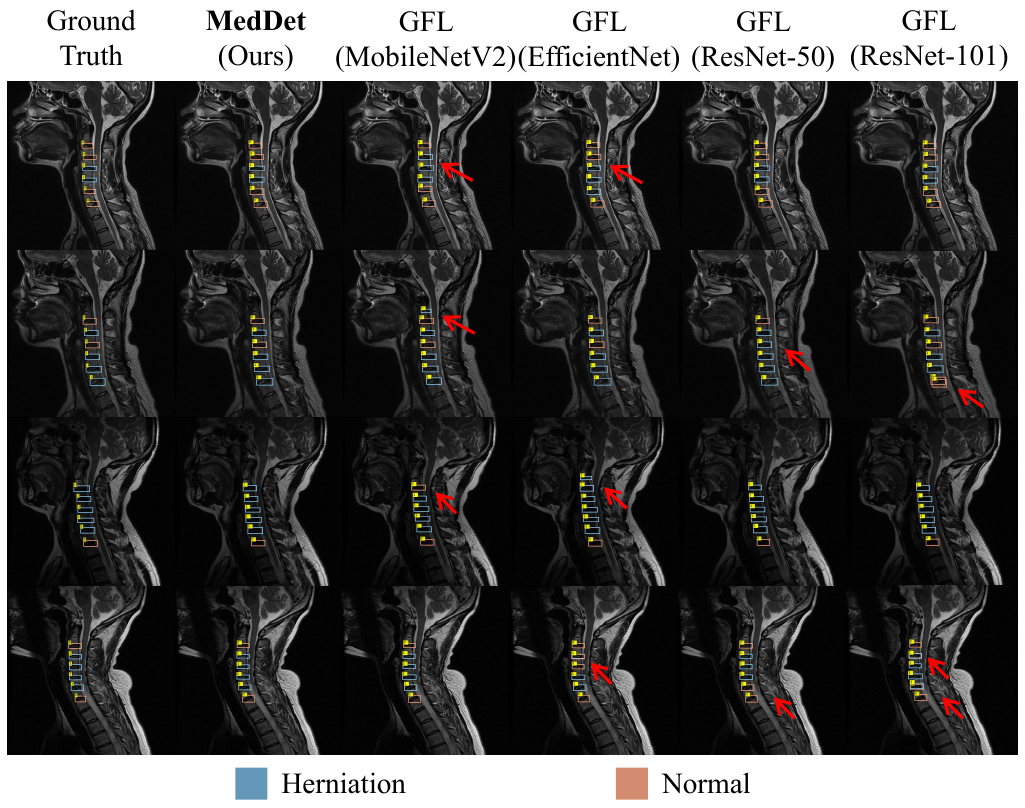}
\caption{The visualization shows that our MedDet surpasses other efficient detection methods, demonstrating its superior ability in accurate CDH detection.}
\label{fig:VISILATION}
\end{figure}

To demonstrate our exceptional performance in CDH detection, we conducted a thorough comparison on CDH-1848 dataset with the Generalized Focal Loss (GFL) \cite{li2020generalized} with efficient backbones such as MobileNetV2 \cite{sandler2018mobilenetv2}, EfficientNet \cite{tan2019efficientnet}, and ShuffleNet \cite{zhang2018shufflenet,ma2018shufflenet} as key baselines. We also compared our method with state-of-the-art knowledge distillation methods in object detection, including MGD \cite{yang2022masked}, LD \cite{zheng2023localization}, and FGD \cite{yang2022focal}. The results in Table \ref{tab:table7} show that our method outperforms these well-established methods, highlighting significant advancements in CDH detection. The visualization in Figure \ref{fig:VISILATION} demonstrates that our method accurately detects CDH compared to other methods. Despite the small size of intervertebral discs in MRI images, our model effectively identifies and diagnoses them, showcasing its strength in detecting small targets.

\begin{table}[h]
\centering
\caption{
The results indicate that our proposed MedDet surpasses other efficient supervised methods and knowledge distillation techniques with the same student network, GFL (MobileNetV2). This demonstrates the superior ability of MedDet to leverage both feature extraction and knowledge transfer, resulting in enhanced performance in CDH detection.}
\label{tab:table7}
\renewcommand{\arraystretch}{1.3}
\resizebox{0.65\linewidth}{!}{%
\begin{tabular}{@{}ccc@{}}   
\toprule
Methods& mAP & Recall \\
GFL (MobileNetV2) \cite{li2020generalized} & 80.8 & \textbf{0.976} \\
GFL (MobileNetV2-1.4) \cite{li2020generalized} & 81.0 & 0.975 \\
GFL (EfficientNet-B0) \cite{li2020generalized} & 80.8 & \textbf{0.976} \\
GFL (EfficientNet-B1) \cite{li2020generalized} & 81.0 & 0.975 \\
GFL (ShuffleNet) \cite{li2020generalized} & 75.0 & 0.964 \\
GFL (ShuffleNet V2) \cite{li2020generalized} & 77.8 & 0.966 \\
\midrule
MGD \cite{yang2022masked}  & 84.2 & 0.969 \\  
LD \cite{zheng2023localization} & 84.7  & 0.970 \\ 
FGD \cite{yang2022focal} & 84.5 & 0.970 \\ 
\textbf{MedDet} (Ours) & \textbf{85.5}  & 0.972 \\ 
\bottomrule
\end{tabular}
}
\end{table}

To demonstrate the exceptional balance between performance and efficiency of our model, we compared it with the teacher models. As shown in Table \ref{tab:table1}, our model achieves an inference speed more than 5 times faster in terms of FPS, along with a 67.8\% reduction in parameters and a 36.9\% reduction in GFLOPs compared to the smallest teacher model, while maintaining comparable performance to teacher models.

\begin{table}[h]
\centering
\caption{The table presents a comparison between our model and the teacher models, demonstrating that our method is significantly lighter and faster while achieving similar performance. This highlights our model's exceptional balance between performance and efficiency.}
\label{tab:table1}
\renewcommand{\arraystretch}{1.3}
\resizebox{\linewidth}{!}{%
\begin{tabular}{@{}cccccc@{}}
\toprule
Methods & mAP & Recall & FPS & Parameters(M) & GFLOPs \\ 
\midrule
GFL (ResNet-152) & \textbf{86.1} & \textbf{0.976} & 2.3 & 66.51 & 92.13 \\ 
GFL (ResNet-101) & 85.8 & 0.974 & 2.5 & 51.24 & 71.20 \\  
GFL (ResNet-50) & 85.6 & 0.970 & 3.9 & 32.32 & 50.21 \\
\midrule
\textbf{MedDet} (Ours) & 85.5 & 0.972 & \textbf{19.8} & \textbf{10.42} & \textbf{31.69} \\ 
\bottomrule
\end{tabular}
}
\vspace{-0.5cm}
\end{table}

\section{Ablation Studies}

To evaluate each component of our proposed architecture, we conducted a thorough ablation study involving nmODE$^{2}$, LWFF, and AATM. The results, presented in Table \ref{tab:table3}, demonstrate that each customization significantly contributes to the overall performance.

\begin{table}[h]
\centering
\caption{The table shows the results of the ablation study on the proposed architecture components including nmODE$^{2}$, LWFF, and AATM, highlighting the significant contributions of each customization to the overall performance.}
\label{tab:table3}
\renewcommand{\arraystretch}{1.3}
\resizebox{0.8\linewidth}{!}{%
\begin{tabular}{@{}ccccc@{}}   

\toprule
        + nmODE$^{2}$ & + LWFF & + AATM & mAP & Recall\\
\midrule
         -  & - & - & 84.2 & 0.968 \\
         $\surd$ & -  & - & 85.0(+0.8) & 0.969 \\
         $\surd$ & $\surd$ & -  & 85.3(+1.1) & 0.971 \\
         $\surd$ & $\surd$ & $\surd$  &\textbf{85.5(+1.3)} & \textbf{0.972} \\

\bottomrule
\end{tabular}
}
\end{table}

We also explore different customizations of nmODE$^{2}$ in various components of the overall architecture. The results, shown in Table \ref{tab:table2}, present different ablations of nmODE$^{2}$ adapted to the student model, GFL (MobileNetV2, KD), with only knowledge distillation. The findings indicate that modifying the backbone and FPN with nmODE$^{2}$ does not lead to significant performance changes, as nmODE$^{2}$ primarily functions in further denoising after feature extraction and lacks the ability to fuse multi-scale features. In contrast, we observe a substantial improvement when nmODE$^{2}$ is used in the detection head, suggesting that nmODE$^{2}$ is more effective at denoising within the detection head.

\begin{table}[h]
\centering
\caption{The table shows the results of different customizations of nmODE$^{2}$ across various components of the overall architecture, highlighting the impact on performance when adapted to the detection head.}
\label{tab:table2}
\renewcommand{\arraystretch}{1.3}
\resizebox{\linewidth}{!}{%
\begin{tabular}{@{}cccccc@{}}   
\toprule
Methods & Backbone & FPN & Head & mAP & Recall \\ 
\midrule
GFL (MobileNetV2, KD) & - & - & - & 84.2 & 0.968\\
\midrule
\multirow{3}{*}{+ nmODE$^{2}$} & $\surd$ & - & - & 83.6 & 0.968 \\
 & - & $\surd$ & - & 84.3 & 0.967 \\
 & - & - & $\surd$ & \textbf{85.0} & \textbf{0.969} \\
\bottomrule
\end{tabular}
}
\end{table}

We also compared different feature alignment and fusion methods. The results in Table \ref{tab:table4} demonstrate that our proposed LWFF achieves the most promising alignment and fusion for knowledge distillation.

\begin{table}[h]
\centering
\caption{The table shows the results of different feature alignment and fusion methods, indicating that our proposed LWFF effectively aligns and fuses features for knowledge distillation.}
\label{tab:table4}
\renewcommand{\arraystretch}{1.3}
\resizebox{0.45\linewidth}{!}{%
\begin{tabular}{@{}ccc@{}}   
\toprule
Methods & mAP & Recall \\ 
\midrule
Sum & 85.1 & 0.969 \\ 
Concatenation & 85.0 & 0.970 \\  
\textbf{LWFF} & \textbf{85.5} & \textbf{0.972} \\ 
\bottomrule
\end{tabular}
}
\end{table}

\vspace{-0.5cm}
\section{Conclusion}

In conclusion, we introduce MedDet, a novel and efficient method for cervical disc herniation (CDH) detection, designed to tackle challenges including noise in MRI images and real-time clinical application. We developed an innovative adversarial auxiliary teacher module (AATM) that incorporates multi-teacher single-student knowledge distillation to enhance the learning capability of the student network. Additionally, we implemented adaptive feature alignment (AFA) and learnable weighted feature fusion (LWFF) to better align the student network’s features with those of the teacher networks. To further improve the model's resilience to MRI noise, we introduced a novel denoising block nmODE$^{2}$ to enhance feature extraction for intervertebral disc imaging. Our comprehensive experiments on the CDH-1848 dataset show that our method achieves up to a \textbf{5\%} improvement in mAP compared to previous methods. Additionally, it achieves over \textbf{5} times faster inference speed and reduces parameters by about \textbf{67.8\%} and FLOPs by \textbf{36.9\%} compared to the teacher model. These results highlight an extraordinary balance between performance and efficiency, marking a significant advancement in CDH detection and showcasing its promising potential for real-world clinical applications.

\section{Acknowledgement}
The work is supported by the Scientific and Technological Innovation Team for Qinghai-Tibetan Plateau Research in Southwest Minzu University (Grant No.2024CXTD20). We would like to appreciate our collaborators for their valuable contributions and support.


\end{document}